\documentclass[aps,pra,twocolumn,groupedaddress,nofootinbib,notitlepage,showpacs,floatfix,superscriptaddress]{revtex4-2}
\usepackage{amsmath}
\usepackage{amsfonts}
\usepackage{graphicx}
\usepackage{dcolumn}
\usepackage{bm}
\usepackage{amsthm}
\usepackage{silence}
\usepackage{times}
\usepackage{color}
\usepackage{xcolor}
\usepackage[colorlinks=true,urlcolor=blue,citecolor=blue]{hyperref}
\usepackage{url}
\usepackage{soul}
\usepackage{graphicx}
\usepackage{amsmath}
\usepackage{appendix}
\usepackage[labelfont=bf,labelformat=empty,labelsep=space,position=auto]{subfig}
\usepackage{setspace}
\usepackage{ragged2e}
\usepackage{xcolor}
\usepackage{amssymb}
\usepackage{latexsym}
\usepackage{times}
\DeclareSubrefFormat{myparens}{#1(#2)}
\usepackage{hyperref}
\DeclareGraphicsExtensions{.png,.eps}
\usepackage{float}
\usepackage{ulem}
\usepackage{appendix}
\usepackage{etoolbox}

\usepackage{xcolor}
\usepackage[urlcolor=blue]{hyperref}      
\hypersetup{
	colorlinks = true,                    
	citecolor = {blue},
	linkcolor = {blue},
}
\begin{document}	
	
	\title{Unconventional photon blockade in a hybrid optomechanical system with an embedded spin-triplet}
	
	\author{Yao Dong}
	\affiliation{School of Physics, Beihang University,100191,Beijing, China}
	
	\author{Jing-jing Wang}
	\affiliation{School of Physics, Beihang University,100191,Beijing, China}
	\author{Guo-Feng Zhang}
	\email{gf1978zhang@buaa.edu.cn}
	\affiliation{School of Physics, Beihang University,100191,Beijing, China}

	\date{\today}
	
	\begin{abstract}
		The research article studies the unconventional photon blockade effect in a hybrid optomechanical system with an embedded spin-triplet state. The interaction between the optomechanical system and the spin state generates new transition paths for the destructive quantum interference of the two-photon excitation state. By analytically solving the Schr\"{o}dinger equation and numerically simulating the master equation, it can be found that the modulated mechanical dissipation is essential for achieving the strong photon blockade in our system. Unlike the conventional cavity optomechanical system, the second-order correlation function $g^{(2)}(0)\simeq 0$ can be obtained with the weak single-photon optomechanical coupling. By adjusting the system parameters, the strong photon blockade and the single-photon resonance can coincide, which indicates the hybrid system has the potential to be a high-quality and efficient single-photon source. Finally, the influence of the thermal noise on photon blockade is investigated. The results show that the second-order correlation function is more robust for the weaker phonon-spin coupling.
		
	\end{abstract}
	
	\maketitle
	
	\section{INTRODUCTION}
	Optomechanical system~\cite{RevModPhys.86.1391,PhysRevLett.121.220404,PhysRevLett.97.237201,Millen_2020,10.1063/1.4930166,PhysRevLett.108.120801}, an attractive platform to explore the light-matter interaction induced by radiation pressure, has achieved significant progress in recent years. Thanks to the great development of nanomechanical fabricating technologies, the optomechanical devices with mass ranging form zettagram to kilogram have been proposed~\cite{Smith_2009,RevModPhys.86.1391}, which can help people understand the quantum-classical boundary via studying macroscopic quantum mechanics, such as macroscopic entanglement~\cite{WOS:000430793000042,PhysRevA.89.014302,PhysRevLett.98.030405,Wang_2019}, ground-state cooling~\cite{https://doi.org/10.1002/lpor.201900120,PhysRevLett.110.153606,PhysRevA.98.023816,PhysRevLett.119.123603,PhysRevA.98.023860,Liu:18}, mechanical squeezing~\cite{PhysRevLett.103.213603,WOS:000360646800047,Xiong:20}, quantum superposition state~\cite{PhysRevLett.116.163602,Xiong:19}, etc. At the same time, the motion of the mechanical resonator has feedback effects on the optic field of the optomechanical system. Considerable research efforts have been devoted to optomechanically induced transparency~\cite{PhysRevA.101.043820,WOS:000356134200001,WOS:000631087100004}, optical amplification~\cite{PhysRevA.95.033803,Jiang:18} and  absorption~\cite{PhysRevA.92.033829,PhysRevA.95.063825},  continuous variable entanglement~\cite{PhysRevLett.110.253601,PhysRevA.103.023525}, fast and slow light~\cite{PhysRevA.92.023846,PhysRevA.107.033507}, photon blockade~\cite{PhysRevLett.107.063601,PhysRevLett.107.063602,PhysRevA.88.023853,WOS:000314104700010,PhysRevA.93.063860,PhysRevA.92.033806,WOS:000535919200035,PhysRevA.99.043818,WOS:001043729600001,PhysRevA.98.013826,PhysRevA.87.013839,PhysRevResearch.5.023148,PhysRevA.102.043705}, etc. In analogy to the famous Coulomb blockade effect~\cite{PhysRevLett.107.216804}, the photon blockade is a nonclassical antibunching effect of quantized  electromagnetic field, which can be used to construct the single-photon source and is of considerable importance for fundamental studies in quantum information processing~\cite{PhysRevLett.114.170802} and quantum computing~\cite{WOS:000995301300003}. Especially after the single-photon induced phonon blockade has been reported~\cite{PhysRevA.99.013804}, the photon blockade has a pivotal role in the future photon-phonon quantum networks.
	
	Photon blockade, in which the excitation of the first photon diminishes the probability of generating the following photon, can be divided into two categories based on physical mechanism. (i) The first one utilizes the anharmonic eigenenergy spectrum from strong nonlinearity~\cite{PhysRevLett.79.1467,PhysRevA.49.R20,PhysRevB.87.235319}, in which the two-photon excitation state is off-resonance. This scheme is referred to as the ``conventional photon blockade (CPB)". Based on the intrinsic nonlinearity of optomechanical system (OMS), the CPB has been theoretically reported in both linearly~\cite{PhysRevLett.107.063601,PhysRevLett.107.063602} and quadratically~\cite{PhysRevA.88.023853,PhysRevA.93.063860} coupled OMS. According to the CPB, the similar phonon blockade has also been explored in the OMS~\cite{PhysRevA.95.053844,WOS:000424061700005,PhysRevA.98.023819,PhysRevA.99.013804,PhysRevA.96.013861}. However, the strong single-photon optomechanical coupling~\cite{PhysRevLett.107.063601} required by CPB experiment in OMS is a major problem. (ii) Another mechanism, known as unconventional photon blockade (UPB) , relied on the destructive quantum interference between different transition paths from the vacuum state to the two-photon excitation state~\cite{PhysRevLett.104.183601,PhysRevA.100.033814,WOS:000314865600005,PhysRevLett.121.043601}. To break the usual limitation of the strong-coupling in the standard OMS, studies begin to emphasize the UPB in the OMS gradually. Related strategies that have been reported include coupling the OMS to an auxiliary cavity~\cite{WOS:001043729600001}, parametric amplification~\cite{WOS:000535919200035}, coupling a $ \mathrm{\Lambda} $-type atom to the cavity mode\cite{PhysRevA.102.043705} and coupling a qubit to the mechanical resonator~\cite{PhysRevA.92.033806}. In all these studies, the presence of additional parts other than the OMS constructs new transition paths, which can facilitate the destructive quantum interference of the two-photon state.
	
	In this paper, we propose an unconventional photon blockade in a hybrid OMS with an embedded spin-triplet state. In our system, a single nitrogen-vacancy (NV) center is embedded in the membrane. According to Ref.~\cite{PhysRevLett.110.156402,PhysRevLett.113.020503}, the transition of the single NV center can be coupled to the mechanical resonant mode via the local strain. Moreover, utilizing Zeeman splitting, we can set the energy level of the spin states. After adiabatically eliminating the excited state, we find there are two necessary factors to obtain the strong photon blockade. First, the coupling between the OMS and the spin-triplet state provides an additional path from the ground state to the two-photon excited state. Second, the modulation of mechanical loss promotes the photon blockade by strengthening the destructive quantum interference. The analytic and numerical calculation of the second-order correlation function indicate that $g^{(2))}(0)\simeq 0$ can occur when the single-photon optomechanical coupling $g\ll\omega_m$. We analytically optimize the parameter condition of the perfect blockade case, which the exact Hamiltonian numerical simulation can verify. To enhance the emission efficiency of the single-photon source, we discuss the strategy to make the single-excitation resonance and the strong photon blockade coincide.  Moreover, we present the method to suppress the negative effect of thermal noise.

	This paper is organized as follows. In Sec.~\ref{section2}, we describe the hybrid spin-optomechanical system and simplify the Hamiltonian utilizing adiabatic elimination under the large detuning approximation. In Sec.~\ref{section3}, we demonstrate the strong photon blockade in the hybrid system by the second-order correlation function. In addition, the optimal parameter relation is analytically derived and numerically verified. A conclusion is given in Sec.~\ref{section4}.

	\section{SYSTEM AND HAMILTONIAN}\label{section2}
	\begin{figure}[tbp]
		\centering
		\subfloat[]{\includegraphics[width=0.95\linewidth]{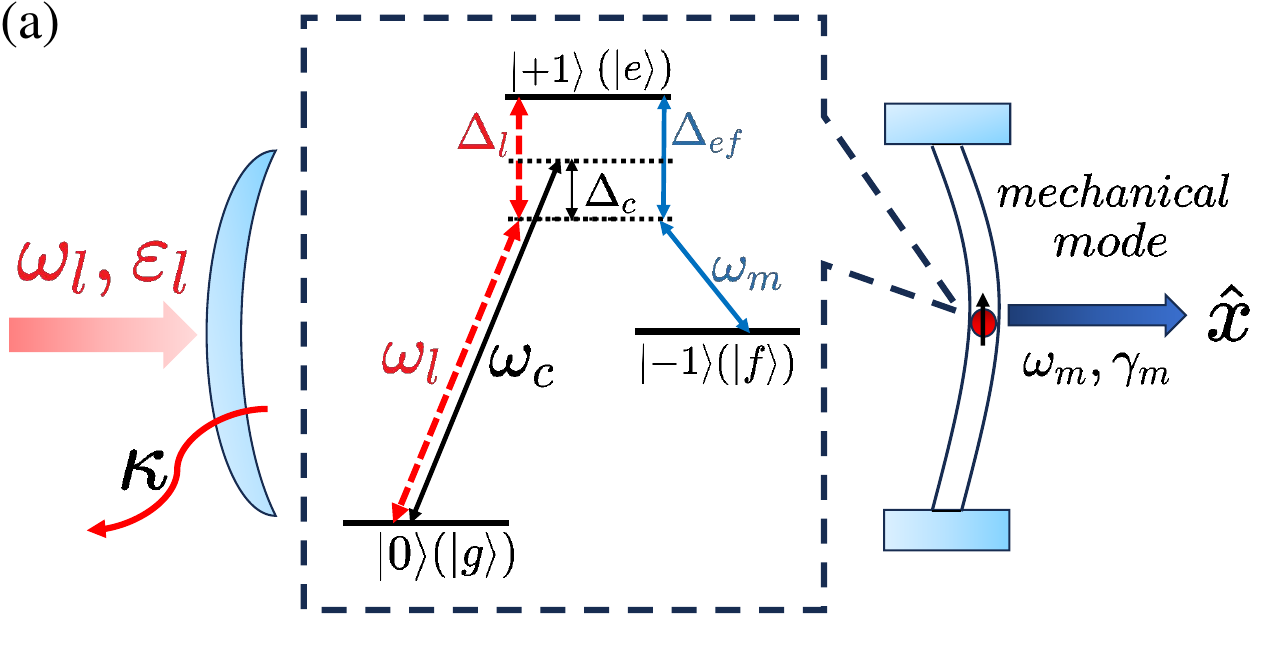}\label{schematic}}\\%
		\centering
		\subfloat[]{\includegraphics[width=0.95\linewidth]{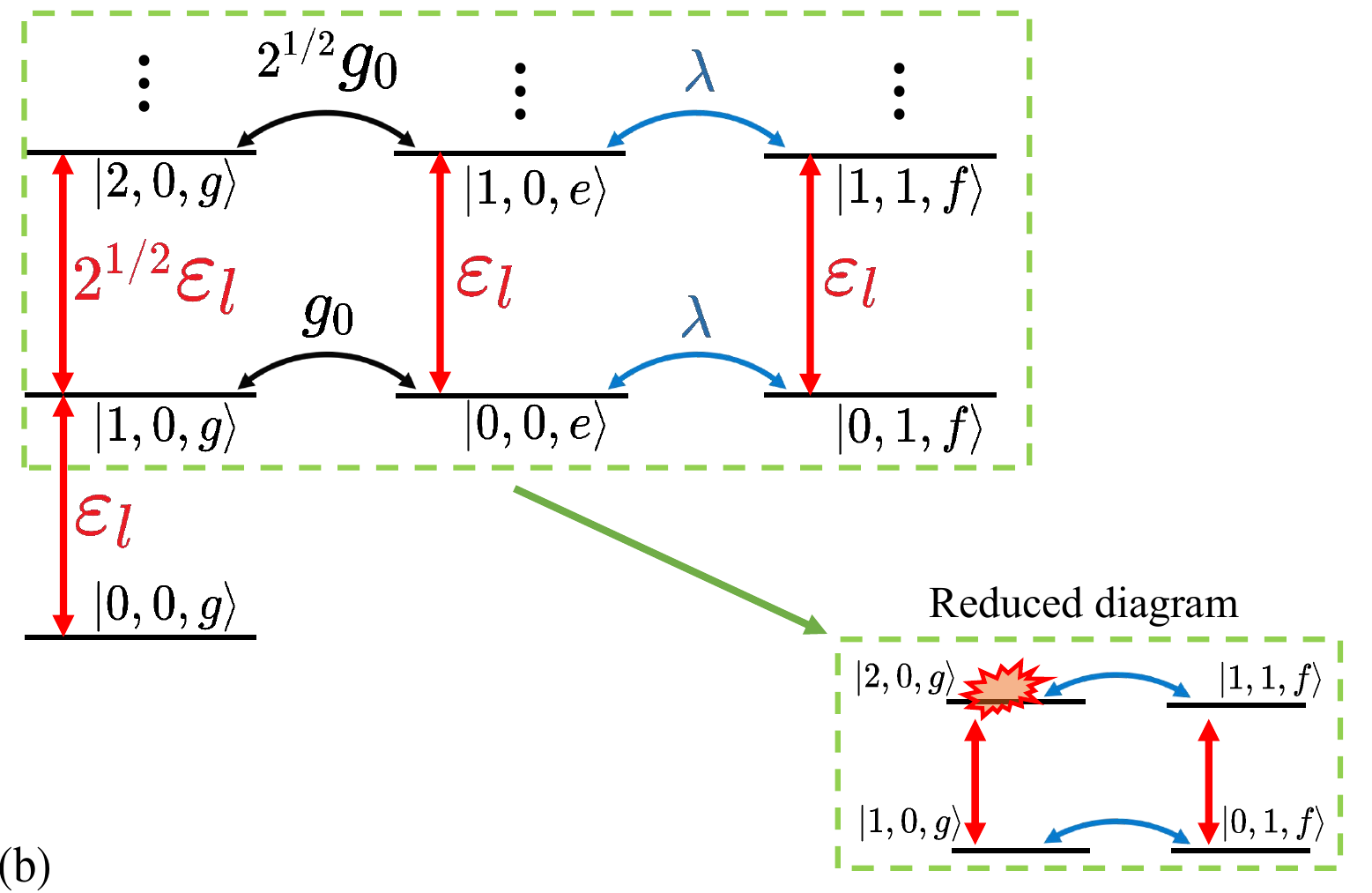}\label{energy_level}}
		
		\caption{\justifying{(a) Schematic illustration of the hybrid optomechanical system including a single NV center spin-triplet state embedded in the mechanical resonator. The NV center includes three spin states $\left\lbrace|m_s=0\rangle, |m_s=+1\rangle,|m_s=-1\rangle\right\rbrace $.~(b) The energy-level diagram within a few-photon subspace, including the photon excitation and interaction between the NV center and the optomechanical system. At its bottom-right corner, the reduced diagram represents the reduced energy level of the approximate Hamiltonian derived via adiabatically eliminating the excited state $|+1\rangle(|e\rangle)$.}}
	\end{figure}
	Let us consider a hybrid optomechanical system where the NV center spin embedded in the membrane as shown in Fig. \subref*{schematic}. The motion of the membrane can change the local strain at the position of the NV center. Then, it will yield a strain-induced electric field that directly couples the membrane (mechanical resonator) to the NV center spin-triplet state~\cite{PhysRevLett.110.156402}. The energy level structure of the NV center depends on the zero field splitting $D_0/2\pi\simeq2.88~\mathrm{GHz}$ and  Zeeman splitting. At the same time, the membrane is linearly coupled to the optical mode via radiation pressure.We also consider the magnetic coupling between the single NV spin and the cavity mode, which can be achieved in a coplanar waveguide (CPW) cavity~\cite{PhysRevB.81.241202,PhysRevA.96.032342,PhysRevB.103.174106}. The system Hamiltonian reads ($\hbar=1$)
	\begin{align}\label{app-eq-H}
		H_1=&\omega_ca^\dagger a+\omega_mb^\dagger b+\omega_e | e \rangle \langle e | +\omega_f | f \rangle \langle f | \notag \\
		&-ga^\dagger a(b+b^\dagger)+g_0(a| e \rangle \langle g |+a^\dagger| g \rangle \langle e |) \notag \\
		&+\lambda (b|e\rangle \langle f |+b^\dagger |f\rangle \langle e |)+H_d,
	\end{align}
	where
	\begin{align}\label{H_d}
		H_d=\varepsilon_l(ae^{i\omega_l t}+a^\dagger e^{-i \omega_l t}).
	\end{align}
	In Eq.~(\ref{app-eq-H}) $a(a^\dagger)$ is the annihilation(creation) operator of the cavity mode with frequency $\omega_c$, $b(b^\dagger)$ is the annihilation(creation) operator of the mechanical resonator with resonance frequency $\omega_m$. We choose the energy of ground state $|g\rangle$ as the zero potential energy point of the spin-triplet system, which is represented by the operators $|l\rangle\langle k|(l,k=g,f,e)$. The single-photon optomechanical coupling strength is $g$. The parameter $g_0$ denotes the coupling strength between the transition $|g\rangle\leftrightarrow |e\rangle $ and the cavity mode, and $\lambda$ denotes the coupling strength between the transition $|e\rangle\leftrightarrow |f\rangle$ and mechanical resonator. In Eq.~(\ref{H_d}), $H_d$ describes the classical laser filed driving with amplitude $\varepsilon_l$ and frequency $\omega_l$.~For the weak driving circumstance, we can deal with this term using the perturbation method.
	
	To simplify the total Hamiltonian, we perform the polaron transform defined by 
	\begin{align}
		V_1=\mathrm{exp}\left[ \dfrac{g}{\omega_m}a^\dagger a(b^\dagger-b)\right].
	\end{align}
	After this transformation, the term of radiation pressure in  Eq.~(\ref{app-eq-H})  will be decoupled. We can get the transformed Hamiltonian
	\begin{align}\label{H_2}
		H_2&=V_1^\dagger H_1 V_1\notag\\
		&=\omega_c a^\dagger a+\omega_m b^\dagger b+\omega_e|e\rangle \langle e|+\omega_f|f\rangle \langle f|\notag\\
		&-\dfrac{g^2}{\omega_m}(a^\dagger a)^2+g_0 ae^{-\frac{g}{\omega_m}(b-b^\dagger)}|e\rangle\langle g|+\lambda b|e\rangle\langle f|\notag\\
		&+\lambda a^\dagger a\dfrac{g}{\omega_m}|e\rangle\langle f|+\varepsilon_l a^\dagger e^{\frac{g}{\omega_m}(b-b^\dagger)} e^{-i\omega_lt}+\mathrm{H.c.},
	\end{align}
	where $g^2/\omega_m$ is the strength of the Kerr-like photon-photon interaction. Considering the fact that the single-photon optomechanical coupling in the practical OMSs is weak, i.e., $g\ll\omega_m$, we limit our system in the weak coupling parameter area and approximately omit exponential factors $\mathrm{exp}[\pm g/\omega_m(b-b^\dagger)]$ and tiny operator $a^\dagger a \frac{g}{\omega_m}$ of Eq.~(\ref{H_2}). In Sec.~\ref{section3} , we will validate the validity of all the approximates via numerical simulation of quantum master equation. In the rotating frame defined by the unitary transformation
	\begin{align}
		V_2=\mathrm{exp}(-i\omega_lta^\dagger a-i\omega_et|e\rangle\langle e|-i\omega_ft|f\rangle\langle f|-i\omega_mtb^\dagger b),
	\end{align}
	the transformed Hamiltonian $H_3=V_2^\dagger H_2 V_2-iV_2^\dagger\dot{V_2}$ is obtained as
	\begin{align}\label{H_3}
		H_3=&\Delta_ca^\dagger a-\dfrac{g^2}{\omega_m}(a^\dagger a)^2+g_0a|e\rangle\langle g|e^{i\Delta_lt}\notag\\
		&+\lambda b|e\rangle\langle f|e^{-i(\omega_m-\omega_{ef})t}
		+\varepsilon_l a+\mathrm{H.c.},
	\end{align}
	where $\Delta_c=\omega_c-\omega_l$, $\Delta_l=\omega_e-\omega_l$, and  $\omega_{ef}=\omega_e-\omega_f$ are corresponding frequency difference.
	
	Under the large detuning condition$(\Delta_l\gg g_0,\lambda)$, we can reduce the freedom of Hamiltonian $H_3$  by  adiabatically eliminating the excited state $|e\rangle$. In addition, we assume $\Delta_l$ and $\omega_{ef}$ satisfy the photon-phonon resonant condition, i.e. $\Delta_l=\omega_{ef}-\omega_m$. Following the adiabatic elimination procedure, we can obtain the reduced Hamiltonian
	\begin{align}\label{H_4}
		H_4=&\Delta_c a^\dagger a-G(a^\dagger a)^2-G_0a^\dagger a|g\rangle\langle g|-\Lambda b^\dagger b|f\rangle\langle f|\notag\\
		&-\sqrt{G_0\Lambda}(a^\dagger b|g\rangle\langle f|+ab^\dagger|f\rangle\langle g|)+\varepsilon_l(a+a^\dagger).
	\end{align}
	For simplify, we have renormalized the system parameters as
	\begin{align}
		G_0=\dfrac{g_0^2}{\Delta_l},\quad\Lambda=\dfrac{\lambda^2}{\Delta_l},\quad G=\dfrac{g^2}{\omega_m}.
	\end{align}
	In Eq.~(\ref{H_4}), $G_0a^\dagger a|g\rangle\langle a|$ and $\Lambda b^\dagger b|f\rangle\langle f|$ are the Stark shifts originating from spin-cavity coupling and spin-mechanics coupling respectively. The interaction term describes the process of one photon creation (annihilation) and one phonon annihilation (creation) as the transition $|g\rangle\leftrightarrow  |f\rangle$. $\sqrt{G_0\Lambda}$ is the effective coupling strength of the resonant interaction between states $|n_a,m_b,g\rangle$ and $|(n-1)_a,(m+1)_b,f\rangle$, where $n_a(m_b)$ denotes the photon (phonon) number of the optical (mechanical) mode and 
	$\left\lbrace g,f\right\rbrace $ denote the spin states. The effective transition processes within few-photon subspace are depicted in Fig.~\subref*{energy_level}. Note that there are two different transition paths from the ground state $|0,0,g\rangle$ to the two-photon excited state $|2,0,g\rangle$, $|0,0,g\rangle\to|1,0,g\rangle\to|2,0,g\rangle$ and $|0,0,g\rangle\to|1,0,g\rangle\to|0,1,f\rangle\to|1,1,f\rangle\to|2,0,g\rangle$. In Sec.~\ref{section3} we will analytically calculate and numerically validate the optimal parameter condition in which strong photon blockade can be realized utilizing the destructive quantum interference of two-photon state $|2,0,g\rangle$.

	\section{UNCONVENTIONAL PHOTON BLOCKADE IN THE HYBRID OPTOMECHANICAL SYSTEM}\label{section3}
	In Sec.~\ref{section2}, we have simplified the hybrid optomechanical system by performing polaron transform and adiabatic elimination under the weak coupling and large detuning conditions. To investigate the photon blockade, here we analytically and numerically calculate the second-order correlation of the optical mode.
	
	\subsection{Analytical solution}
	
	We will obtain the analytical solution of second-order correlation function by approximately solving the Schr\"{o}dinger equation in a truncated subspace like the general research of photon blockade. We consider both optical and mechanical dissipation and phenomenologically add the imaginary terms into the reduced Hamiltonian~(\ref{H_4}). Under the weak optomechanical coupling condition, the effective non-Hermitian Hamiltonian $H_{\mathrm{eff}}$ is written as (see Appendix~\ref{appendix_A})
	\begin{align}\label{H_eff}
		H_{\mathrm{eff}}=H_4-i\dfrac{\kappa}{2}a^\dagger a-i\dfrac{\gamma_m}{2}\left(\dfrac{g}{\omega_m}a^\dagger a\right)^2
		-i\dfrac{\gamma_m}{2}b^\dagger b,
	\end{align}
	where $\kappa$ and $\gamma_m$ are the decay rates of the optical cavity and the mechanical oscillator, respectively. We assume that the average lifetime of the spin excited state is much longer than that of photon and phonon, i.e.,$\gamma\ll \kappa,\gamma_m$. So the decay rate of spin-triplet system in the effective non-Hermitian Hamiltonian is not considered. Under the limit of the weak driving, the population of high level can be neglected. Then the time-dependent quantum state $|\psi(t)\rangle$ of the hybrid system can be expressed as
	\begin{align}\label{state}
		|\psi(t)\rangle=&C_{00g}(t)|0,0,g\rangle+C_{10g}(t)|1,0,g\rangle+C_{01f}(t)|0,1,f\rangle\notag\\
		&+C_{20g}(t)|2,0,g\rangle+C_{11f}(t)|1,1,f\rangle,
	\end{align}
	where $C_{nmg}(C_{nmf})$ is the probability amplitude of the quantum state $|n,m,g\rangle(|n,m,f\rangle)$.~Although $|0,0,g\rangle$, $|1,0,g\rangle$, $|0,1,f\rangle$, $|2,0,g\rangle$ and $|1,1,f\rangle $ are a set of bases in the mechanical displacement representation, they can be considered orthogonal due to the weak coupling limitation. 
	
	Substituting  the effective Hamiltonian and the truncated quantum state into the Schr\"{o}dinger equation $i\mathrm{d}|\psi(t)\rangle/\mathrm{d}t=H_{\mathrm{eff}}|\psi(t)\rangle$, we can obtain a set of differential equations for the probability amplitudes,
	\begin{align}\label{differential}
		&i\frac{\partial C_{10g}}{\partial t} =\Delta_1 C_{10g}-\sqrt{G_0\Lambda}C_{01f}+\varepsilon_l C_{00g}+\sqrt{2}\varepsilon _l C_{20g},\notag\\
		&i\frac{\partial C_{01f}}{\partial t} =\Delta_f C_{01f}-\sqrt{G_0\Lambda}C_{10g}+\varepsilon_lC_{11f},\notag\\
		&i\frac{\partial C_{20g}}{\partial t} =2\Delta_2 C_{20g}-\sqrt{2G_0\Lambda}C_{11f}+\sqrt{2}\varepsilon _lC_{10g},\\
		&i\frac{\partial C_{11f}}{\partial t} =(\Delta_3+\Delta_f)C_{11f}-\sqrt{2G_0\Lambda}C_{20g}+\varepsilon _l C_{01f},\notag
	\end{align}
	where
	\begin{align}
		&\Delta_1=\Delta_c-\frac{i}{2}\kappa-\frac{i}{2}\gamma_m\left(\frac{g}{\omega_m}\right)^2-G-G_0,\notag\\
		&\Delta_2=\Delta_c-2G-G_0-\frac{i}{2}\kappa-i\gamma_m\left(\frac{g}{\omega_m}\right)^2,\notag\\
		&\Delta_3=\Delta_c-G-\frac{i}{2}\kappa-\frac{i}{2}\gamma_m\left(\frac{g}{\omega_m}\right)^2,\notag\\
		&\Delta_f=-\Lambda-\frac{i}{2}\gamma_m.
	\end{align}
	In the weak driving region $\varepsilon_l\ll\kappa$, there exists the fact $\left\lbrace C_{20g},C_{11f}\right\rbrace\ll\left\lbrace C_{10g},C_{01f} \right\rbrace \ll C_{00g}  $, so we can set $C_{00g}\simeq 1$.~By neglecting the high-order small terms $\left\lbrace \varepsilon_lC_{20g},\varepsilon_lC_{11f}\right\rbrace $ and setting the time derivative be zero, the probability amplitudes of steady-state can be approximately derived as
	\begin{align}\label{steady_solution}
		&C_{10g}=\dfrac{-\Delta_f\varepsilon_l}{\Delta_1\Delta_f-G_0\Lambda},\notag\\
		&C_{01f}=\dfrac{-\sqrt{G_0\Lambda}\varepsilon_l}{\Delta_1\Delta_f-G_0\Lambda},\notag\\
		&C_{20g}=\dfrac{\varepsilon_l^2(\Delta_3\Delta_f+\Delta_f^2+G_0\Lambda)}{\sqrt{2}(\Delta_1\Delta_f-G_0\Lambda)(\Delta_2\Delta_3+\Delta_2\Delta_f G_0\Lambda)},\notag\\
		&C_{11f}=\frac{\varepsilon_l^2\sqrt{G_0\Lambda}(\Delta_2+\Delta_f)}{(\Delta_1\Delta_f+G_0\Lambda)(\Delta_2\Delta_3+\Delta_2\Delta_f G_0\Lambda)}.
	\end{align}
	
	Next we characterize the photon blockade effect by the equal-time second-order correlation function
	\begin{align}\label{g20}
		g^{(2)}(0)=\frac{\langle a^\dagger a^\dagger a a\rangle}{\langle a^\dagger a\rangle^2},
	\end{align}
	which represents the probability of detecting two photons simultaneously.  $g^{(2)}(0)>1$ indicates the photon-induced tunneling effect, i.e. the excitation of a photon will enhance the probability of exciting the subsequent photon. On the other hand, $g^{(2)}(0)<1$ indicates the completely opposite photon blockade effect that the first excited photon will suppress the followed one. Specially, the situation $g^{(2)}(0)\rightarrow 0$ means the perfect photon blockade. Utilizing the analytical solution of the probability amplitudes in Eq.~(\ref{steady_solution}), the second-order function of the steady state can be written as (see Appendix~\ref{appendix_B})
	\begin{align}\label{g20_analytical}
		g^{(2)}(0)=\dfrac{2|C_{20g}|^2}{(|C_{10g}|^2+|C_{11f}|^2+2|C_{20g}|^2)^2}\simeq \dfrac{2|C_{20g}|^2}{|C_{10g}|^4},
	\end{align}
	where we use the fact $|C_{20g}|,|C_{11f}|\ll|C_{10g}|$ under the condition of weak driving. To find the optimal parameter relation to achieve strong UPB, we set the probability amplitude of the two-photon state to equal zero, i.e.$C_{20g}=0$. Subsequently the optimal parameters can be directly derived as (see details in Appendix~\ref{appendix_B})
	\begin{align}\label{optimal}
		&\Delta_c=\dfrac{\kappa+\gamma_m}{\gamma_m}\Lambda+G+\Lambda,\notag\\
		&G_0=(\Lambda+\dfrac{\gamma_m^2}{4\Lambda})\dfrac{\kappa+\gamma_m}{\gamma_m}.
	\end{align}
	Due to the multi-photon excitation $(n\ge3)$ and the approximations used in the previous derivation, the zero-delay second-order correlation function $g^{(2)}(0)$ cannot be suppressed entirely to zero, although in the analytical optimal parameter relation. Subsequent numerical simulation indicates that $g^{(2)}(0)$ can be suppressed to $4\times10^{-3}$ in our system. The analytical solution of $g^{(2)}(0)$ in Eq.~(\ref{g20_analytical}) will be verified via simulating the Lindblad master equation of the accuracy Hamiltonian~(\ref{app-eq-H}).
	
	\subsection{Numerical simulation by the master equation}
	Next, we calculate the exact numerical solution of the two-order correlation function and the dynamics of the hybrid system by the method of the Lindblad master equation. For convenience, we preform a rotating transformation defined by $V'$ on the initial Hamiltonian. The transformation is defined by
	\begin{align}
		V'=\mathrm{exp}(-i\omega_lta^\dagger a-i\omega_lt|e\rangle\langle e|-i\omega_lt|f\rangle\langle f|),
	\end{align}
	which doesn't contain any approximations. Then the accuracy Hamiltonian can be rewritten as
	\begin{align}
		H'=&\Delta_c a^\dagger a+\omega_mb^\dagger b+\Delta_l |e\rangle \langle e| -(\omega_{ef}-\Delta_l)|f\rangle \langle f|\notag\\
		&-ga^\dagger ab+g_0a|e\rangle\langle g|+\lambda b^\dagger|f\rangle\langle e|+\varepsilon_la+\mathrm{H.c.},
	\end{align}
	where $\Delta_c=\omega_c-\omega_l$ is the detuning between the cavity mode and the driving field. Then the Lindblad master equation of the hybrid system is given by
	
	\begin{align}\label{master_equation}
		\dfrac{\partial\rho}{\partial t}=&-i[H',\rho]+\kappa\mathcal{D}[a]\rho+\gamma_m(\bar{n}_0+1)\mathcal{D}[b]\rho+\gamma_m \bar{n}_0\mathcal{D}[b^\dagger]\rho\notag\\
		& + \dfrac{1}{2T_2}\mathcal{D}[|e\rangle\langle e|-|f\rangle\langle f|]\rho+\gamma \mathcal{D}[|g\rangle \langle f|]\rho\notag\\
		&-\gamma(|e\rangle\langle e|\rho-|g\rangle\langle e|\rho|e\rangle\langle g|-|f\rangle\langle e|\rho|e\rangle\langle f|+\rho|e\rangle\langle e|),
	\end{align}
	where $\mathcal{D}[o]\rho=o\rho o^\dagger-(o^\dagger o\rho+\rho o^\dagger o)/2$ denotes the Liouvillian in Lindblad form, $\bar{n}_0=[\mathrm{exp}(\hbar\omega_m/k_BT)-1]^{-1}$ is the mean thermal phonon occupation number at temperature $T$ and $\gamma$ is the decay rate of the spin system. Here we assume $\gamma$ to be the same for the transitions of $|e\rangle\to |f\rangle$, $|e\rangle\to |g\rangle$ and $|f\rangle\to|g\rangle $. Moreover, the NV center would be subjected to a single dephasing $T_2^{-1}$. We ignored single spin relaxation as $T_1$ can be several minutes at low temperature~\cite{PhysRevLett.108.197601}.~After a long enough time, the hybrid optomechanical system will evolve to its steady state $\rho_s$.~And the steady-state second-order correlation function is given by
	\begin{align}
		g^{(2)}(0)=\dfrac{\mathrm{Tr}(a^\dagger a^\dagger a a\rho_s)}{[\mathrm{Tr}(a^\dagger a\rho_s)]^2}.
	\end{align}
	
	During the numerical simulation, we truncated the system dimension to $5\otimes5\otimes3$ (5 Fock states for photon, 5 Fock states for phonon and 3 spin states for the single NV center) in Fig.~\subref*{fig2a} $\sim$ Fig.~\subref*{mechanical_loss} and $5\otimes6\otimes3$ in Fig.~\subref*{thermal}.~The principle for truncating at the given number of the Fock state is that the relative fluctuation rates of the second-order correlation function are less than $10^{-5}$ if we continue truncating the system to a larger Hilbert space. Such a tiny variation means that the given truncated system can reflect the actual situation to a considerable extent.
	\subsection{Induce additional mechanical damping}\label{addition_damping}
	Generally, the optical cavity loss rate $\kappa$ exceeds the intrinsic mechanical damping $\gamma_m$ by many orders of magnitude, which causes the optimal parameters to be divergent in Eq.~(\ref{optimal}). In order to have $g^{(2)}(0)\ll 1$ we consider that the mechanical oscillator is coupled to an ancillary cavity of decay $\kappa_a$ which is driven at the mechanical red sideband $\Delta_a=-\omega_m$ such as to play an effective low-temperature thermal reservoir. When the optomechanical coupling $G$ between the ancillary cavity and mechanical modes is much smaller than $\kappa_a$, we can adiabatically eliminate the ancillary cavity modes and induce an additional damping $\gamma_{\mathrm{opt}}$ to the oscillator. Then the effective mechanical damping and occupation number are~\cite{PhysRevLett.99.093902,PhysRevA.91.061803,PhysRevA.92.013852} (for details see Appendix~\ref{appendix_C})
	\begin{align}\label{effeciv damping}
		\Gamma_m=\gamma_m+\gamma_{\mathrm{opt}},\quad\bar{n}=\dfrac{\gamma_m\bar{n}_0+\gamma_{\mathrm{opt}}\bar{n}_{\mathrm{opt}}}{\gamma_m+\gamma_{\mathrm{opt}}}.
	\end{align}
	Here, $\gamma_m$ and $\bar{n}_0$ are the intrinsic linewidth and environment thermal occupation number of the bare mechanical oscillator, and $\bar{n}_{\mathrm{opt}}=(\kappa_a/4\omega_m)^2$ is the quantum limit of sideband cooling\cite{PhysRevLett.99.093902,PhysRevLett.99.093901}.~In the resolved-sideband regime, we can control $\gamma_{\mathrm{opt}} = 4|G|^2/\kappa_a$  by setting the effective optomechanical coupling $G$, e.g. driving the ancillary cavity with different laser power. Moreover, the "dressed" mechanical oscillator and the bare one have the same form of the Langevin equation~\cite{PhysRevA.91.061803} and the  master equation~\cite{PhysRevA.95.023827}, only the damping rate $\gamma_m\to\Gamma_m$ and the thermal phonon number $\bar{n}_0\to\bar{n}$ are modified. 
	
	We can note that the role of laser cooling is to provide a small effective phonon occupation number $\bar{n}\ll\bar{n}_0$ while simultaneously increasing the effective mechanical damping $\Gamma_m\gg\gamma_m$ by the same factor. It has been confirmed that laser cooling can facilitate the observation of a quantum feature such as sub-Poissonian phonon statistics~\cite{PhysRevA.91.061803}. Next we will show its similar effect to photon statistical properties.
	\subsection{Photon statistical properties}
	
\begin{figure}[tbp]
		\centering
		\subfloat[]{\includegraphics[height=1.6in]{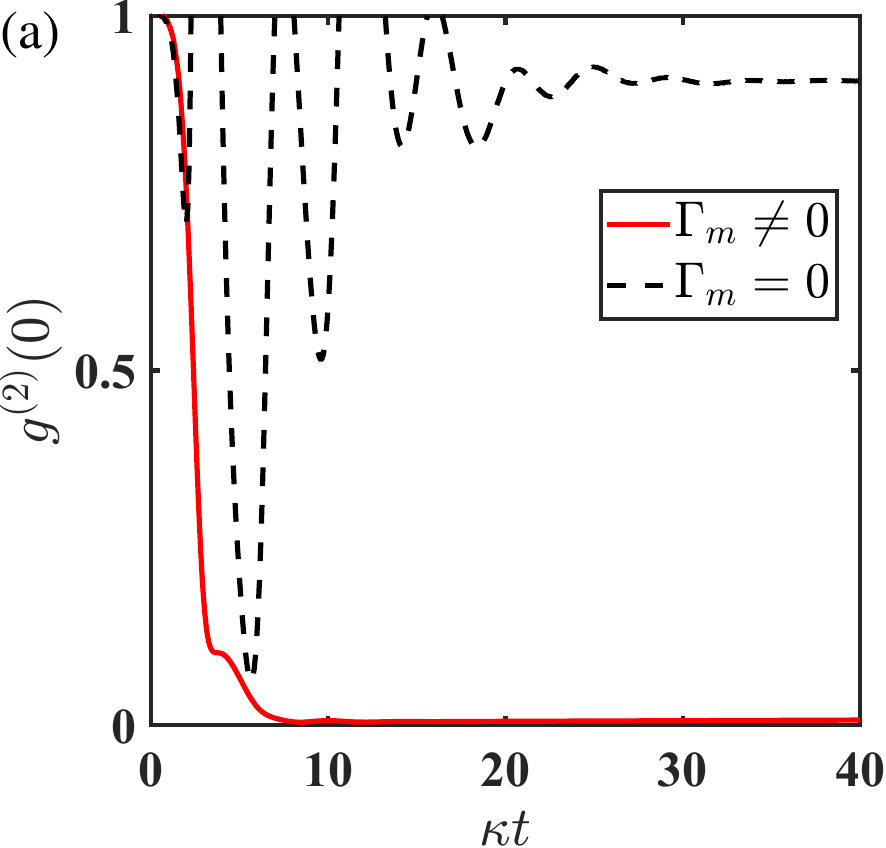}\hspace{1mm}\label{fig2a}}%
		\centering
		\subfloat[]{\includegraphics[height=1.7in]{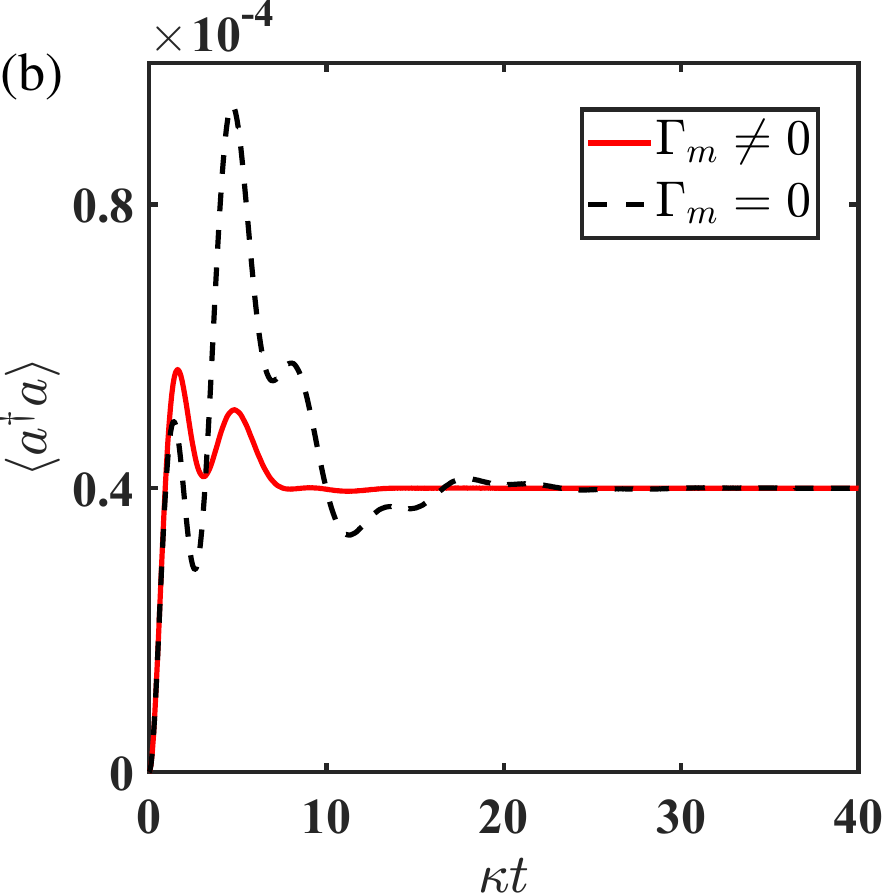}\label{fig2b}}
		
		\caption{\justifying{The dynamical evolution of (a) the zero-delay second-order correlation function $g^{(2)}(0)$ and (b) mean intracavity photon number $\langle a^\dagger a\rangle$ with or without modulation of mechanical dissipation. The system parameters are chosen as $\kappa=2\pi$ MHz, $\omega_m/\kappa=100$, $\Delta_l=\omega_{m}$, $g/\omega_m=0.03$, $G_0=2\kappa$, $\Lambda=0.5\kappa$, $\varepsilon_l=0.01\kappa$, $T_2=1$ms, $\bar{n}=0$, $\Gamma_m=\kappa$ (solid red line) or $\Gamma_m=0$ (dashed black line).}}\label{Fig2}
	\end{figure}

	We first consider a particular case in which the dissipation of the mechanical oscillator is modulated to the same as the optical cavity, i.e.~$\Gamma_m=\kappa$.
	According to the optimal parameter relations in Eq.~(\ref{optimal}), the perfect photon blockade effect occurs at the optimal detuning $\Delta_c=3\Lambda+G$ and the coupling $G_0=2\Lambda+\kappa^2/2\Lambda$. To preliminary validate the analytical optimal parameter relations and the presence of the steady-state, we solve the dynamical evolution of the correlation function $g^{(2)}(0)$ and mean intracavity photon number $\langle a^\dagger a\rangle$ via the numerical simulation from Eq.~(\ref{master_equation}), as shown in Fig.~\ref{Fig2}. We can find that the hybrid system can evolve to its steady-state whether or not the dissipation modulation exists. The steady value of the second-order correlation function $g^{(2)}(0)$ can be obtained respectively at $\kappa t\approx 15$ or  $\kappa t\approx 30$ when the dissipation modulation exists or not, which indicates that additional mechanical damping has resulted in shorter relaxation time of the system. One can see from Fig.~\subref*{fig2a} that perfect steady-state photon blockade phenomenon ($g^{(2)}(0)\simeq 0$) can be achieved under the optimal parameter relations with mechanical dissipation modulation (solid red line), while only weak steady-state photon antibunching effect ($g^{(2)}(0)\simeq 0.9$) occurs without mechanical dissipation modulation (dashed black line), i.e. $\Gamma_m\to 0$.

	Fig.~\subref*{fig2b} can further verify the accuracy of the steady-state probability amplitudes in Eq.~(\ref{steady_solution}). Under weak driving condition, the steady value of mean intracavity photon number $\langle a^\dagger a\rangle\simeq \left | C_{10g} \right | ^2$. Substituting optimal parameters into the probability amplitudes of the single-photon state $|1,0,g\rangle$ in Eq. (\ref{steady_solution}), we can obtain $\langle a^\dagger a\rangle\simeq 0.4\varepsilon_l^2$ when the mechanical damping $\gamma_m=\kappa$ or $\gamma_m=0$, which agrees well with the numerical simulations in Fig.~\subref*{fig2b}. Different from Ref.~\cite{WOS:000535919200035}, the stronger photon blockade effect doesn't necessarily correspond to the lower mean photon number in our system. For a high-Q mechanical resonator the intrinsic damping can be discarded as $\kappa \gg \gamma_m$, but we retain a finite damping $\gamma_m=10^{-4}\kappa$ for the situation without dissipation modulation to simulate the exact dynamical evolution. Moreover, we first focus on the case at zero temperature and the effect of the thermal phonon occupation number $\bar{n}$ on photon blockade will be discussed in the following subsection.

\begin{figure}[tbp]
		\centering
		\subfloat[]{\includegraphics[angle=0,height=1.62in]{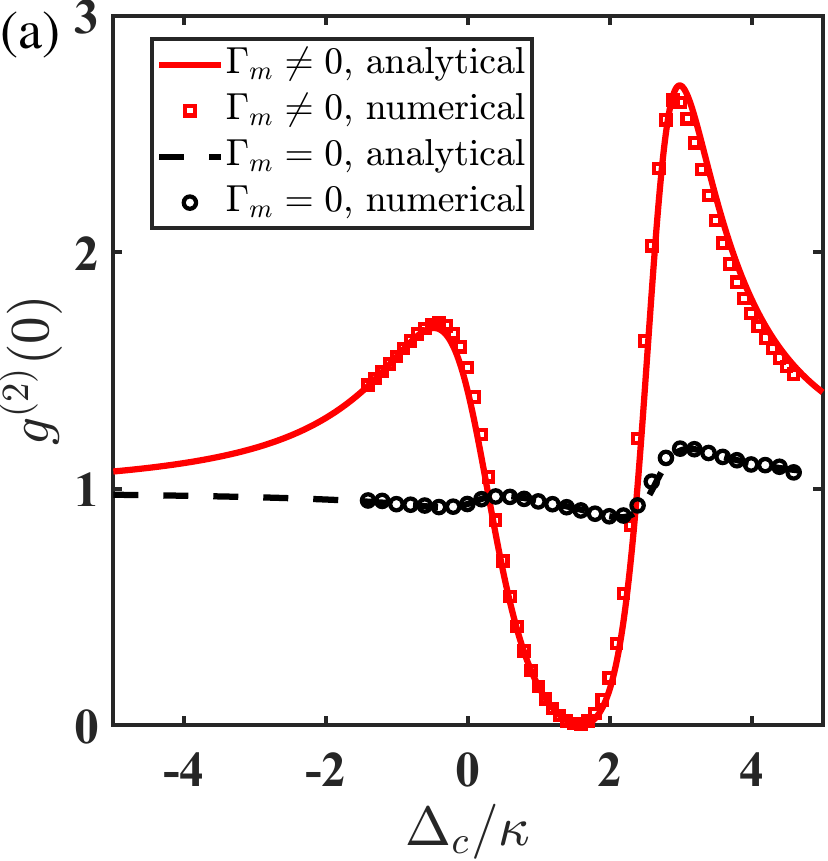}\label{fig3a}}\hspace{1mm}%
		\centering
		\subfloat[]{\includegraphics[angle=0,height=1.72in]{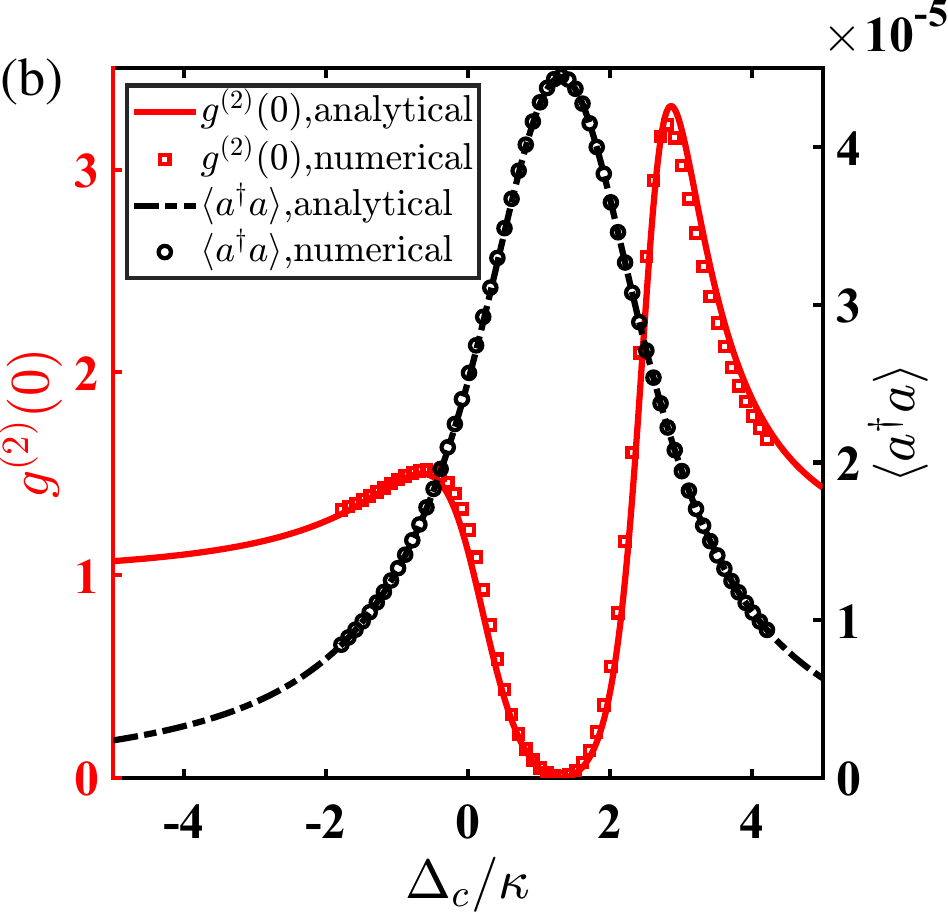}\label{fig3b}}
		
		\caption{\justifying{(a) Zero-delay second-order correlation function $g^{(2)}(0)$ versus the cavity-pumping detuning $\Delta_c$ for the modulation of mechanical dissipation existing or not. The other parameters are the same as in Fig.~\ref{Fig2}. (b) Second-order correlation function $g^{(2)}(0)$ and mean intracavity photon number $\langle a^\dagger a\rangle$ of the high-quality and efficient single-photon source versus the detuning $\Delta_c$. Parameters are taken as $G_0=2.04\kappa$, $\Lambda=0.408\kappa$, and the other parameters are the same as in Fig.~\ref{Fig2}.}}\label{Fig3}
	\end{figure}

	In Fig.~\subref*{fig3a}, we plot the variation of second-order correlation function $g^{(2)}(0)$ with the cavity-pumping detuning $\Delta_c$ via analytical and numerical simulation. In the case of dissipation modulation, we can see that the strong photon blockade ($g^{(2)}(0)\simeq0$) occurs at the optimal detuning given in Eq.~(\ref{optimal}). In addition, the analytical steady value of $g^{(2)}(0)$ (solid red line) and the numerical solution (red square) are well consistent in the detuning region where antibunching effect appears. The second-order correlation function of phonon at the optimal detuning is numerically calculated to be as low as 0.013, which suggests that our system has the potential to be a high-quality single-photon and single-phonon source at the same time. In Fig. \subref*{fig3a} there are two peaks of $g^{(2)}(0)>1$, which indicates the process of two-photon resonance. 
	Without dissipation modulation, we can only see the fragile photon blockade phenomenon. From Eq.~(\ref{steady_solution}), the necessity of mechanical dissipation modulation to achieve strong photon blockade in the hybrid system can be confirmed. When $\Gamma_m=0$ the probability amplitudes of steady-state in Eq.~(\ref{state}) can be similarly derived as 
	\begin{align}
		C_{20g}=\dfrac{\varepsilon_l^2\Lambda(\Delta_c-G-\Lambda-\frac{i}{2}\kappa)}{\sqrt{2}(\Delta_1\Lambda+G_0\Lambda)(\Delta_2\Delta_3-\Delta_2\Lambda-G_0\Lambda)}.
	\end{align}
	Obviously, there is no real solution to the equation $|C_{20g}|=0$, i.e., $\Delta_c-G-\Lambda-\frac{i}{2}\kappa=0$, which indicates the perfect UPB cannot occur without the modified mechanical damping. One can also see that the analytical result (dash-dot black line) agrees well with the numerical solution (black circles).
	\begin{figure*}[tbp]
		\centering
		\subfloat[]{\includegraphics[width=0.33\linewidth]{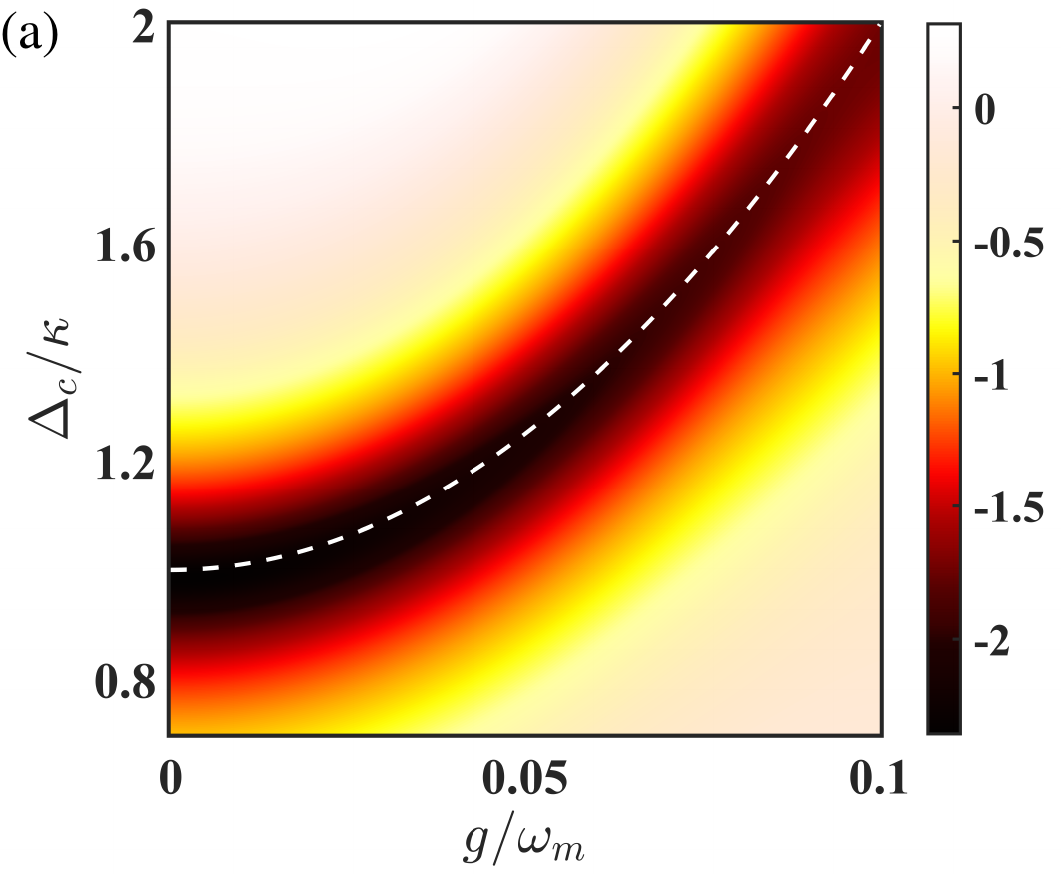}\label{4a}}
		\centering
		\subfloat[]{\includegraphics[width=0.33\linewidth]{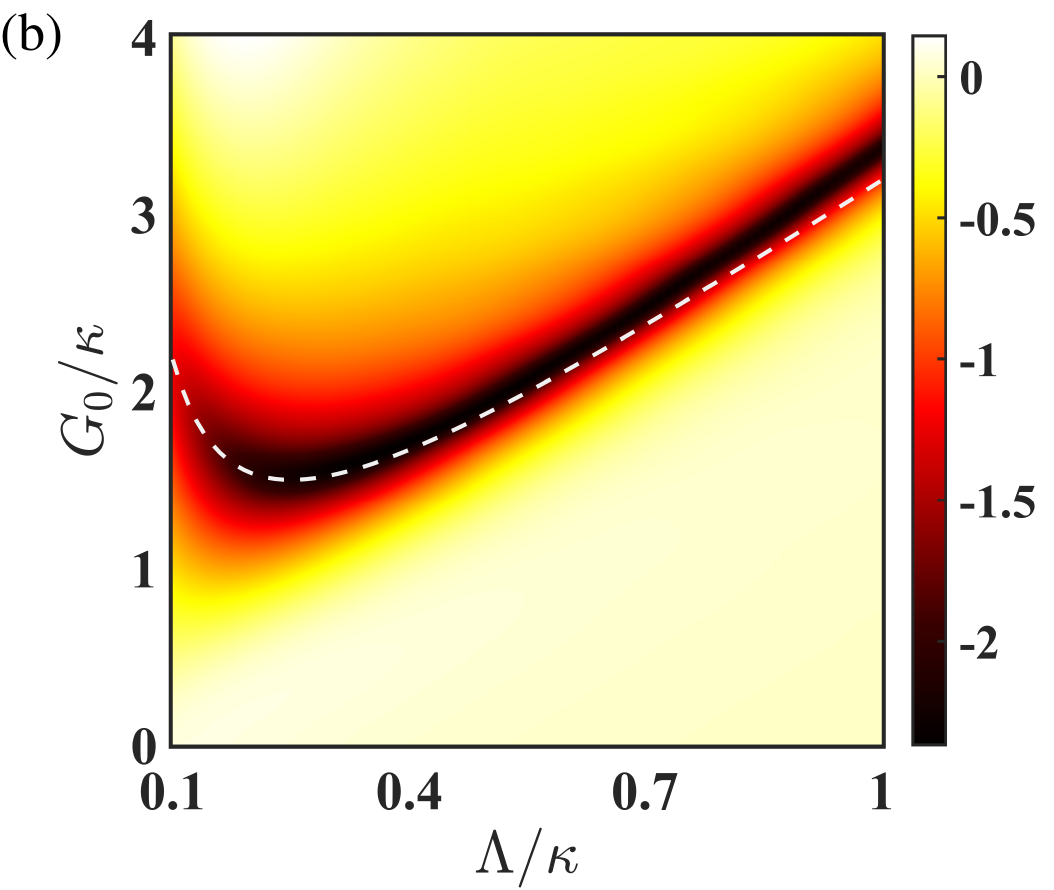}\label{4b}}\hspace{-2mm}
		\centering
		\subfloat[]{\includegraphics[width=0.29\linewidth]{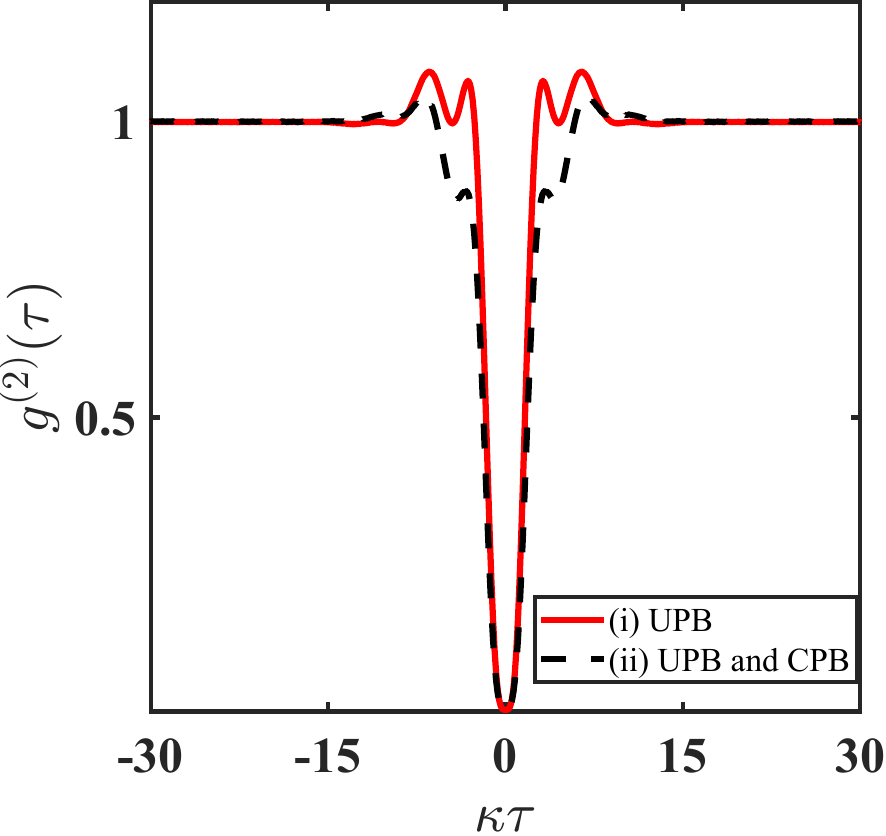}\label{4c}}
		\caption{\justifying{(a) Equal-time second-order correlation function $\log_{10} g^{(2)}(0)$ versus the cavity-pimping detuning $\Delta_C$ and optomechanical coupling $g$. Parameters are taken as $\omega_m=100\kappa$ , $\varepsilon_l=0.01\kappa$, $G_0=1.5\kappa$, $\Lambda=0.25\kappa$, and the mechanical dissipation is modulated to $\Gamma_m=0.5\kappa$. (b) Equal-time second-order correlation function $\log_{10}g^{(2)}(0)$ as a function of the renormalized coupling $G_0$ and $\Lambda$. We set $g=3\kappa$ and the other parameters are the same as in (a). (c) The delayed second-order correlation $g^{(2)}(\tau)$ versus the time-delay $\kappa\tau$. For cases (i) and (ii), the parameters are the same as in Fig.~\ref{Fig2} or Fig.~\protect\subref*{fig3b}, respectively. }}
	\end{figure*}
	
	According to the Eq.~(\ref{optimal}), we can find that the detuning location of the perfect photon blockade can be changed via modifying the optomechanical coupling strength ($G=g^2/\omega_m$), phonon-spin coupling strength ($\Lambda=\lambda^2/\Delta_f$) and mechanical dissipation $\Gamma_m$. To achieve a high-quality and efficient single-photon source, it requires that the perfect photon blockade phenomenon occurs with a high single-photon occupancy, and we have derived the corresponding parameter relation. Maximizing the single-photon state occupying probability  $|C_{10g}|^2$, the cavity-pumping detuning is given by
	\begin{align}\label{max_photon number}
		\Delta_c=G+\dfrac{\gamma_m^2}{4\Lambda^2+\gamma_m^2}G_0,
	\end{align}
	which is the optimal relation of the single-excitation resonance from ground state $|0,0,g\rangle$ to the eigenstate of subspace $\left\lbrace |1,0,g\rangle,|0,1,f\rangle\right\rbrace $. Due to the modulation of the embedded spin-triplet state system to the anharmonic energy-level, it is possible to manipulate the position of the single-photon resonance without changing the optomechanical coupling strength $g$, which is different from Ref.~\cite{WOS:000535919200035}. In Fig.~\subref*{fig3b}, we analytically calculate and numerically simulate the correlation function $g^{(2)}(0)$ and mean intracavity photon number $\langle a^\dagger a\rangle$ with different detuning $\Delta_c$, respectively.
	Combining Eq.~(\ref{optimal}) and Eq.~(\ref{max_photon number}), it can be seen that the prefect photon blockade and the high-efficiency photon emission will occur at the same location. The mean intracavity photon number of the high-quality and efficient single-photon source is almost $4.5\times10^{-5}$ when the driving amplitude $\varepsilon_l=0.01\kappa$, which indicates that the efficiency of single-photon emission is approximately $280$ per second for $\kappa=2\pi\,  \mathrm{MHz}$. Moreover, the analytical results also agree with the numerical simulations for correlation function $g^{(2)}(0)$ and mean photon number $\langle a^\dagger a\rangle$ in Fig.~\subref*{fig3b}.
	
	Next, we turn to discuss the effect of parameter fluctuation on the correlation function and discuss the parameter range in which the approximate analytical derivation is valid. Fig.~\subref*{4a} displays the correlation function $g^{(2)}(0)$ as a function of single-photon optomechanical coupling strength $g$ and cavity-pumping detuning $\Delta_c$ via numerical simulation. It demonstrates that the location of the perfect photon blockade moves to a larger red-detuning with the enhancement of optomechanical coupling. However, for relatively large optomechanical coupling strength $g$, the exponential factors $\mathrm{exp}[\pm g/\omega_m(b-b^\dagger)]$ and tiny operator $\frac{g}{\omega_m}a^\dagger a$ in Eq.~(\ref{H_2}) cannot be safely omitted, which causes the weaker photon blockade effect in the optimal condition. We show the analytical optimal detuning location in Eq.~(\ref{optimal}) by dashed white line, which has the identical trend with the central region of the strong blockade. Like Ref.~\cite{PhysRevA.102.043705}, the strength of optomechanical coupling does not affect the setting of other optimal parameters. It is noteworthy that the strong photon blockade effect can be achieved in the weak optomechanical coupling region $g\ll\omega_m$, which breaks the constraint of strong coupling in the standard cavity optomechanical system.
	
	To discuss the enhancement of the photon blockade due to the coupling between the optomechanical system and the spin-triplet state, Fig.~\subref*{4b} exhibits the correlation function $g^{(2}(0)$ as a function of the renormalized coupling strength $G_0$ and $\Lambda$ via numerical simulation. Similar to Fig.~\subref*{4a}, the dashed white line represents the analytical optimal relation of coupling strength in Eq.~(\ref{optimal}). Overall, the trend of numerical simulation is in accordance with the analytical optimal parameter relations for a strong photon blockade. We find that on the left side of Fig.~\subref*{4b}, the antibunching effect becomes weaker ($g^{(2)}(0)\simeq 0.06$) with the decrease of $\Lambda$ in the optimal parameters condition. The reason is that when the difference of coupling strength between $|n,m,g\rangle \leftrightarrow  |n-1,m,e\rangle$ (denoted by $g_0$) and $|n-1,m,e\rangle\leftrightarrow  |n-1,m+1,f\rangle$ (denoted by $\lambda$) is significant, the population of middle excited state $|e\rangle$ will increase. Then the transition path $|1,0,e\rangle\to |1,1,f\rangle\to |1,0,f\rangle\to |2,0,f\rangle$ that does not exist in the effective reduced energy-levels can cause the extra two-photon population. With the increase of $G_0$ and $\Lambda$ on the right side of Fig.~\subref*{4b}, the central region that perfect photon blockade occurs deviates from the analytical optima relation gradually. This is because for relatively large coupling strength $\Lambda$, the interaction term $\lambda a^\dagger ag/\omega_m(|e\rangle\langle f|+|f\rangle\langle e|)$ in Eq.~(\ref{H_2}) is not negligible and it will move the parameter region of the perfect photon blockade.

	To further characterize the statistical property of the single-photon source, we calculated the delayed second-order correlation function
	\begin{align}\label{g2tau}
		g^{(2)}(\tau)=\lim_{t \to \infty} \dfrac{\langle a^\dagger (t)a^\dagger (t+\tau)a(t+\tau)a(t)\rangle }{\langle a^\dagger(t)a(t)\rangle\langle a^\dagger (t+\tau)a(t+\tau)\rangle},
	\end{align}
	which is proportional to the probability of detecting one photon at time $t$ and detecting the subsequent photon at time $t+\tau$. For the steady state $\rho_s$ of the system, the delayed second-order correlation function can be equivalently defined as
	\begin{align}\label{g2tau}
		g^{(2)}(\tau)=\dfrac{\mathrm{Tr}[a^\dagger aU(\tau)a\rho_sa^\dagger U^\dagger(\tau)]}{[\mathrm{Tr}(a^\dagger a\rho_s)]^2},
	\end{align}
	where $U(\tau)$ is the time evolution operator of the open system.~Because including dissipation and decoherence processes, $U(\tau)$ is a superoperator defined by $U(\tau)\bullet U^\dagger(\tau)=e^{\mathcal{L}\tau}\bullet$, where $\mathcal{L}$ denotes the master equation $\dot{\rho}=\mathcal{L}\rho$ in Eq.~(\ref{master_equation}).~In Fig.~\subref*{4c}, we can see that the value of $g^{(2)}(\tau)$ is always larger than $g^{(2)}(0)$. Like other systems without the gain part in Ref.~\cite{PhysRevA.90.023849,PhysRevA.96.053810,PhysRevA.92.023838}, $g^{(2)}(0)$ finally reaches 1 with the increase of the time delay, so it describes the standard Poisson photon statistics. Moreover, for case (i) (solid red line), $g^{(2)}(\tau)$ exhibits significant oscillation when only UPB occurs. The first photon emitted at time $t$ and the second photon emitted after a time delayed $\tau\approx\pm 3.2/\kappa$ or $\tau\approx\pm 6.4/\kappa$ prefer to arrive at the detector together. As for case (ii) (dashed black line), when UPB and CPB coincide like Fig.~\subref*{fig3b}, the oscillation of $g^{(2)}(\tau)$ can be suppressed, which means a lower time resolution is required for observation~\cite{WEI2023106202}.

	In the previous discussion, we investigated the occurrence of the perfect single-photon blockade with the fixed modulated mechanical dissipation $\Gamma_m$.~In this case, one must flexibly modulate the coupling strength $g_0$ or $\lambda$  to the specific optimal values. Based on the progress in manipulating phonons~\cite{PhysRevLett.116.183602,WOS:000322592000016} and oscillator $\mathcal{PT}$  symmetry~\cite{PhysRevLett.113.053604,PhysRevA.92.013852,PhysRevA.95.023827,PhysRevA.100.063846,PhysRevA.107.033522}, the mechanical damping and gain rates are controllable in a considerable range. Compared to the tunable coupling between the optomechanical and the spin system, dissipation modulating can be experimentally implemented more conveniently. So in this subsection, we will exhibit the optimal mechanical dissipation versus different coupling strengths and analyze the effect of the actual thermal noise on photon blockade.
	
\begin{figure}[tbp]
		\centering
		\hspace{2mm}
		\subfloat[]{\includegraphics[width=0.85\linewidth]{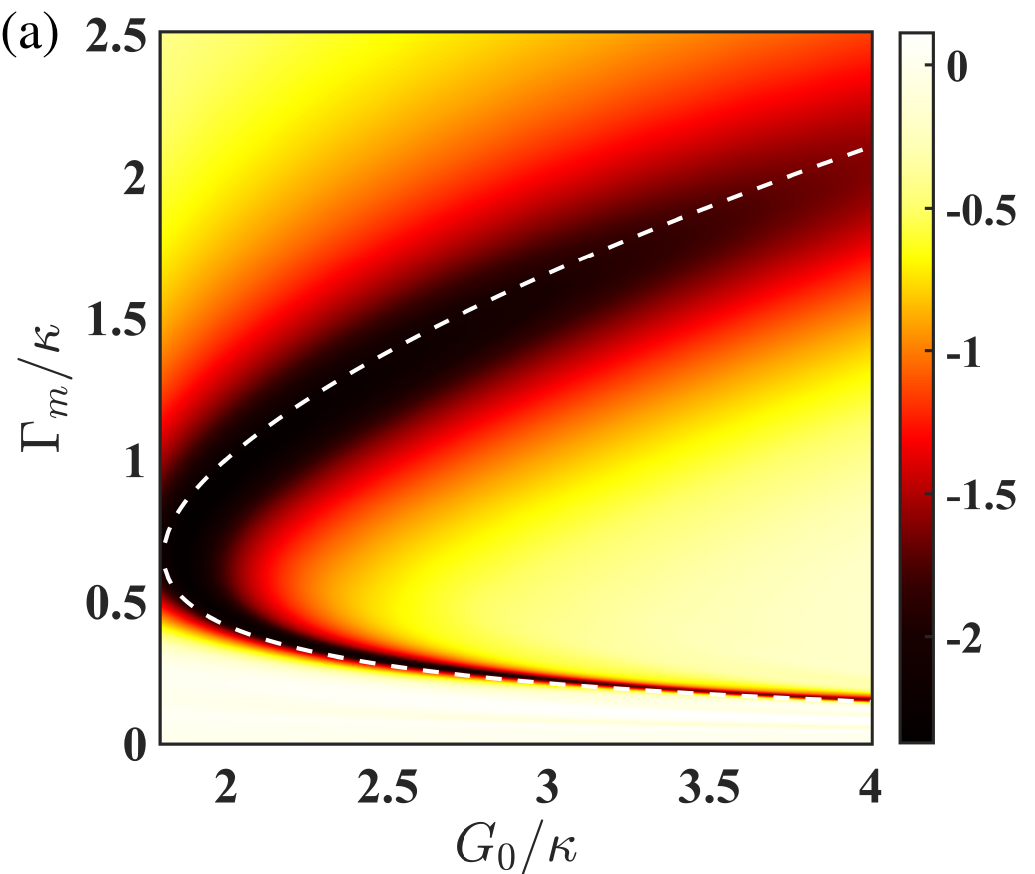}\label{mechanical_loss}}\\%
		\centering
		\hspace{-4mm}
		\subfloat[]{\includegraphics[width=0.78\linewidth]{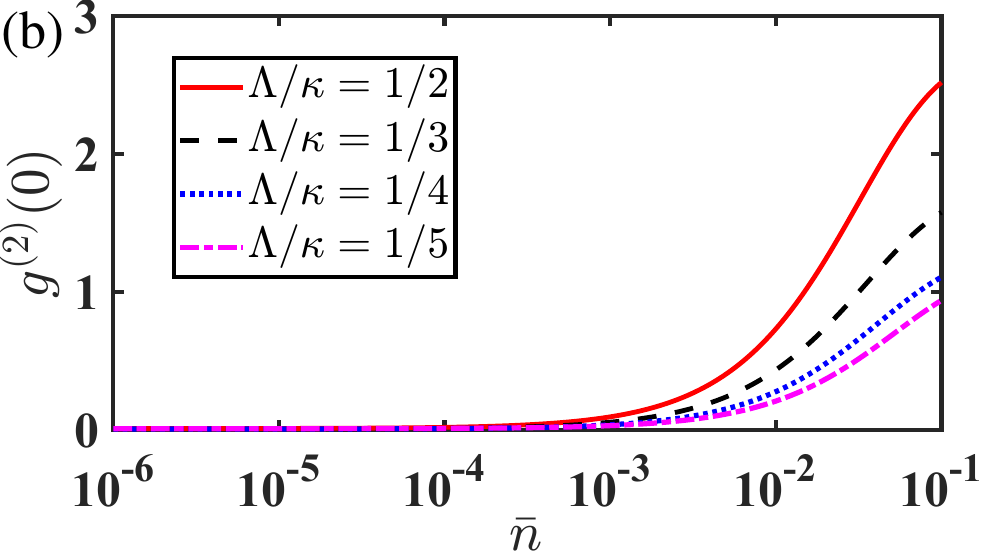}\label{thermal}}
		
		\caption{\justifying{(a) The equal-time second-order correlation function $\log_{10}(g^{(2)}(0))$ versus the modulated mechanical damping $\gamma_m$ and the renormalized coupling $G_0$ with optimal pumping-cavity detuning. We set $\Lambda=\frac{1}{2}\kappa$ and the other parameters are the same as in Fig.~\ref{Fig2}.~(b) Second-order correlation $g^{(2)}(0)$ versus the thermal phonon occupation number $\bar{n}$ with $\Gamma_m=\frac{1}{2}\kappa$. }}
	\end{figure}

	Fig.~\subref*{mechanical_loss} demonstrates the second-order correlation function $\log_{10}(g^{(2)}(0))$ as a function of normalized coupling $G_0$ and modulated mechanical loss $\Gamma_m$. The optimal relation in Eq. (\ref{optimal}) can be written as a cubic equation about modulated mechanical dissipation,
	\begin{align}\label{Eq24}
		\dfrac{1}{4\Lambda}\Gamma_m^3+\dfrac{\kappa}{4\Lambda}\Gamma_m^2+(\Lambda-G_0)\Gamma_m+\kappa\Lambda=0,
	\end{align}
	whose real solutions are represented by the dashed white line in Fig.~\subref*{mechanical_loss}.~We can see that the prefect photon blockade can appear at two different parameter regions, which are consistent with the two real solutions of the optimal mechanical damping $\Gamma_m$ calculated analytically. For the larger modulated one, correlation function $g^{(2)}(0)$ has better robustness to parameter fluctuation. With the increase of $\Gamma_m$, the antibunching effect of the emitted photon is gradually suppressed. The reason is that a larger $\Gamma_m$ will strengthen the decay process from $|1,1,f\rangle$ to $|1,0,f\rangle$, which is a dark state of the reduced Hamiltonian in Eq.~(\ref{H_4}) without the driving term. Then  state $|1,0,f\rangle$ will be driven to $|2,0,f\rangle$. Hence, the population of the two-photon excitation state cannot be eliminated.
	
	To demonstrate the negative effect of the mean thermal phonon number on photon blockade, we numerically simulate $g^{(2)}(0)$ as a function of $\bar{n}$ at optimal parameters in Fig.~\subref*{thermal}. It shows that $g^{(2)}(0)$ increases monotonically with thermal phonon number $\bar{n}$. Under our system parameters, $\bar{n}$ needs to be cooled to $10^{-2}$ or lower so that obvious antibunching effect can be observed. For a beam of dimensions $(l,w,t)=(4.3,0.1,0.1)\mu m$, the resonance frequency $\omega_{m}$ can be calculated by the Euler-Bernoulli theory~\cite{Ref82}, i.e. $\omega_m=(4.73/l)^2\sqrt{EI/\rho A}~\sim 2\pi\times100$ MHz. Assuming an environment temperature 10 mK in a dilution refrigerator~\cite{PhysRevLett.125.153602}, $\bar{n}_0$ is about 1.6. Drawing support from the advanced intracavity-squeezed cooling technology proposed by Yong-Chun Liu et al.~\cite{https://doi.org/10.1002/lpor.201900120}, the minimum final phonon number is $n^{\mathrm{min}}_f\approx\dfrac{2\bar{n}_0}{Q_\mathrm{m}}+\sqrt{\dfrac{\bar{n}_0}{Q_\mathrm{m}}}$, where $Q_\mathrm{m}$ is the quality factor of the beam. Single-crystal diamond mechanical with high Q in excess of $10^6$ have been fabricated~\cite{WOS:000335220900002}, which means we can cool $\bar{n}$ below $1.3\times10^{-3}$; such a low phonon occupancy number is sufficient to observe the strong photon blockade in our system.
	
	It is worth noting that $g^{(2)}(0)$ has the better robustness when coupling $\Lambda$ is smaller. To explain this feature, we assume that the optical mode and spin system are initially in the vacuum state $|0\rangle$ and ground state $|g\rangle$, respectively, while the mechanical mode is in the thermal state at temperature $T$. Then the initial system can by described by  $\rho=(1-p) {\textstyle \sum_{m\ge 0}} p^m|0,m,g\rangle\langle 0,m,g|$, where $p=\mathrm{exp}(-\hbar\omega_m/k_BT)$. So the population of two-photon state cannot be completely suppressed due to the new transition paths $|0,m,g\rangle\to|1,m,g\rangle\to|2,m,g\rangle$ and $|0,m,g\rangle\to|1,m,g\rangle\to|0,m+1,f\rangle\to|1,m+1,f\rangle\to|2,m,g\rangle$. It is easy to find that the larger $\Lambda$ results in a stronger effective coupling strength $\sqrt{G_0\Lambda}$, which can cause the above paths to be more prominent and hinder the photon blockade. 
	
	\section{CONCLUSION}\label{section4}
	
	In conclusion, we have explored the UPB effect in a hybrid optomechanical system with an embedded spin-triplet state. By analytically solving the Schr\"{o}dinger equation of effective Hamiltonian, we obtain the parameter relation of the perfect photon blockade and calculate the second-order correlation function $g^{(2)}(0)$. Then all the analytical derivations and results are verified via numerically simulating the Lindblad master equation of the accuracy Hamiltonian. Notably, when the modulated mechanical dissipation and the coupling of spin-triplet state fulfill the derived analytical optimal parameter relation, a strong photon blockade can occur in the weak optomechanical coupling region $g\ll\omega_{m}$, which breaks the strong-coupling constraint. The existence of the spin-triplet state can create new transition paths for the destructive interference of the two-photon excitation state. To eliminate the two-photon occupation, we find that the mechanical dissipation needs to be modulated to the order of the optical decay rate $\kappa$. Lastly we have investigated the influence of the mechanical thermal noise on the photon blockade. The higher mechanical frequency $\omega_m$ and cooling temperature are beneficial for the antibunching effect. We also explain why the weaker phonon-spin coupling $\lambda$ should be chosen based on optimal parameter relations to suppress the negative effect of the mechanical thermal noise. The modulated mechanical gain rate can also meet the optimal relation of the perfect blockade, and we will expand the hybrid system to non-Hermitian in the future. Our work could lead to achieve single-photon source and photon-phonon networks utilizing the weak coupling optomechanical system.
	\section*{Conflict of Interest}
	The authors declare no conflict of interest.
	\section*{Data availability statement}
	The data that support the findings of this study are available from the corresponding author upon reasonable request.
	
	\begin{acknowledgments}
		This work was supported by NSFC under grants NOS. 12074027.
	\end{acknowledgments}
	
	\appendix
	\section{Effective Non-Hermitian Hamiltonian}\label{appendix_A}
	During the derivation of the reduced Hamiltonian $H_4$ in Eq.~(\ref{H_4}), we have performed the rotating-frame and the polaron transformation. In the rotating picture, it has the effect of shifting the excited state down in energy; so this transformation doesn't impact the decays in the effective Non-Hermitian Hamiltonian of Eq.~(\ref{H_eff}). While in the mechanical displacement representation defined by $V_1=\exp[g/\omega_m a^\dagger a(b^\dagger-b)]$, the annihilation operator of the phonon $b$ transforms to $b+g/\omega_m a^\dagger a$. Thus the effective Non-Hermitian Hamiltonian becomes
	\begin{align}
		H_{\mathrm{eff}}=H_4-i\frac{\kappa}{2}a^\dagger a-i\frac{\gamma_m}{2}\left(b^\dagger+\frac{g}{\omega_m}a^\dagger a\right)\left(b+\frac{g}{\omega_m}a^\dagger a\right).
	\end{align}
	For a weak optomechanical coupling strength $g$, we omit the term $a^\dagger ab^\dagger$ and $a^\dagger ab$ due to $\frac{\gamma_m}{2}\frac{g}{\omega_m}\ll\omega_m$. So we can get Eq.~(\ref{H_eff}) in the main text.
	\section{Analytical Steady State Solution}\label{appendix_B}
	When the system evolve to its steady-state $|\psi_s\rangle$ depending on the probability amplitudes in Eq.~(\ref{steady_solution}), the equal-time second-order correlation can be written as
	\begin{align}
		g^{(2)}(0)=\dfrac{\langle \psi_s |a^\dagger a^\dagger a a|\psi_s \rangle}{\left(\langle \psi_s|a^\dagger a|\psi_s \rangle\right)^2}.
	\end{align}
	For the numerator, only the two-photon state $|2,0,g\rangle$ can contribute a non-zero term $2|C_{20g}|^2$ in the truncated Hilbert space. The denominator of $g^{(2)}(0)$ can be calculated utilizing the square of the mean photon number, i.e. $|C_{10g}|^2+|C_{11f}|^2+|C_{20g}|^2$. Under the weak driving ($\varepsilon_l\ll\kappa$), due to $\left\lbrace
	|C_{11f}|^2,|C_{20g}|^2\right\rbrace$ are many orders of magnitude smaller than $|C_{10g}|^2$, we omit them and reach to Eq.~(\ref{g20_analytical}).
	
	The optimal condition for $g^{(2)}(0)\to 0$ corresponds $C_{20g}=0$, i.e.
	\begin{align}
		\Delta_f(\Delta_3+\Delta_f)+G_0\Lambda=0,
	\end{align}
	where $\Delta_3=\Delta_c-G-\frac{i}{2}\kappa-\frac{i}{2}\gamma_m\left(\frac{g}{\omega_m}\right)^2$ and $\Delta_f=-\Lambda-\frac{i}{2}\gamma_m$. Solving this equation produces the optimal parameter relations
	\begin{align}\label{B3}
		&\Delta_c=\dfrac{\kappa+\gamma_m+2\left(\dfrac{g}{\omega_m}\right)^2
			\gamma_m}{\gamma_m}\Lambda+G+\Lambda,\notag\\
		&G_0=\left(\Lambda+\dfrac{\gamma_m^2}{4\Lambda}\right)\dfrac{\kappa+\gamma_m+2\left(\dfrac{g}{\omega_m}\right)^2
			\gamma_m}{\gamma_m}.
	\end{align}
	Weak single-photon optomechanical coupling condition $g\ll\omega_{m}$ causes $(g/\omega_m)^2\ll1$. After approximately neglecting this tiny term, Eq.~(\ref{B3}) is exactly the same as Eq.~(\ref{optimal}) in the main text.
	\section{Additional Mechanical Damping}\label{appendix_C}
	As the statement in Sec.~\ref{addition_damping}, we consider linearly couple the mechanical oscillator to an auxiliary optical cavity with linewidth $\kappa_a$. After the standard linearization procedure, the Hamiltonian can be written as
	\begin{align}
		H=-\Delta_a a^\dagger a+\omega_m b^\dagger b+G(a+a^\dagger)(b+b^\dagger),
	\end{align}
	where $\Delta_a=\Delta_l-\Delta_a$ is the laser detuning from the cavity resonance and $G$ is the modified optomechanical coupling strength. We assume the red-sideband resonance condition $\Delta_a=-\omega_m$ and $G$ to be real without loss of generality.Subsequently, we have the linearized quantum Langevin equation
	\begin{align}\label{C2}
		&\dot{a}=\left(i\Delta_a-\dfrac{\kappa_a}{2}\right)a-iG(b+b^\dagger)-\sqrt{\kappa_a}a_{in}\notag\\
		&\dot{b}=\left(-i\omega_m-\dfrac{\gamma_m}{2}\right)-iG(a+a^\dagger)-\sqrt{\gamma_m}b_{in}.
	\end{align}
	We move Eq.~(\ref{C2}) into another rotating frame by introducing the slowly moving operators, i.e. $a=\tilde{a}e^{-i\Delta_a t}$ and $b=\tilde{b}e^{-i\omega_m t}$. In the condition of resolved sideband and weak coupling ($\omega_m\gg\kappa_a\gg G$), the scattering processes described by $ab$ and $a^\dagger b^\dagger$ are off-resonance
	and therefore suppressed. Invoking the RWA we can obtain
	\begin{align}
		&\dot{a}=-\dfrac{\kappa_a}{2}a-iGb-\sqrt{\kappa_a}a_{in}\notag\\
		&\dot{b}=-\dfrac{\gamma_m}{2}b-iGa-\sqrt{\gamma_m}b_{in}.
	\end{align}
	In order to derive the effective damping of the mechanical mode, we neglect the input-noise terms for simplicity. In the case of $\omega_m\gg\kappa_a\gg G\gg\gamma_m$, the formal solution of the auxiliary cavity operator $a$ at the long time scale $t\gg1/\kappa_a$ can be expressed as
	\begin{align}\label{C4}
		a(t)=-iGe^{-\frac{\kappa_a}{2}t}\int_{0}^{t}b(t')e^{\frac{\kappa_a}{2}t'}dt'.
	\end{align}
	Then, we adiabatically eliminate the auxiliary cavity modes. As the evolution of $b$ is much slower than $a$, we can set $b(t')\approx b(t)$ and take it out of the integral in Eq.~(\ref{C4}). Evaluating the integral directly, we obtain
	\begin{align}
		&a(t)=-\dfrac{2iG}{\kappa_a}b(t)\notag\\
		&\dot{b}=-\dfrac{1}{2}(\gamma_m+\dfrac{4G^2}{\kappa_a})b,
	\end{align}
	i.e. an addition damping $4G^2/\kappa_a$ is induced to the phonon. For the thermal noise, it can be considered to come from the surrounding thermal environment and the cooling light field whose thermal phonon occupancy are $\bar{n}_0$ and $\bar{n}_{\mathrm{opt}}$, respectively. $\bar{n}_0$ depend on the environmental temperature, i.e. $\bar{n}_0=[\mathrm{exp}(\hbar\omega_m/k_BT)-1]^{-1}$ and $\bar{n}_{\mathrm{opt}}$ is given by the balance expression utilizing the Fermi's golden rule
	\begin{align}
		\dfrac{\bar{n}_{\mathrm{opt}}+1}{\bar{n}_{\mathrm{opt}}}=\dfrac{S_{FF}(+\omega_m)}{-\omega_m}\Rightarrow\bar{n}_{\mathrm{opt}}=\left(\dfrac{\kappa_a}{4\omega_{m}}\right)^2,
	\end{align}
	where $S_{FF}(\omega)$ is the spectrum of the optical force. The effective occupation number $\bar{n}$ follows from the thermal balance between the environment and the cooling light field:
	\begin{align}
		\bar{n}=\dfrac{\gamma_m\bar{n}_0+\gamma_{\mathrm{opt}}\bar{n}_{\mathrm{opt}}}{\gamma_m+\gamma_{\mathrm{opt}}}.
	\end{align}
	The effective mechanical damping and occupation number of the "dressed" mechanical oscillator can also be derived from the master equation's self-consistency~\cite{PhysRevA.95.023827}.

	\bibliography{Primary_manuscript}

	
	
	

\end{document}